\documentclass{aastex}
\usepackage{spr-astr-addons}
\usepackage{url}

\RequirePackage{color}

\begin{document}

\title{The Stokes Phase Portraits of Descattered Pulse Profiles of a Few Pulsars}
\slugcomment{Not to appear in Nonlearned J., 45.}
\shorttitle{stokes Phase Portraits of Descattered Pulse Profiles}
\shortauthors{Abdujappar Rusul et al.}

\author{Abdujappar Rusul\altaffilmark{1}}  \and \author{Ali Esamdin\altaffilmark{2}}
\and
\author{Xiao-Ping Zheng\altaffilmark{1}} \and \author{Jian-Ping Yuan\altaffilmark{2}}



\altaffiltext{1}{The Institute of Astrophysics, Huazhong Normal University, Wuhan 430079, China.}
\altaffiltext{2}{Xinjiang Astronomical Observatory, Chinese Academy of Science, Urumqi 830011, China.}

\begin{abstract}
The observing signals from pulsar are always influenced by the interstellar medium (ISM) scattering. In the lower frequency observation, the intensity profiles are broadened and the plane of polarization angle (PPA) curves are flattened by the scattering effect of the ISM. So before we analyze the scattered signal, we should take a proper approach to clear scattering effect from it. Observing data and simulation have shown that the Stokes phase portraits $I$-$U$, $I$-$Q$ and $Q$-$U$ are also distorted by the ISM scattering. In this paper, a simulation is held to demonstrate a scattering and a descattering of the Stokes phase portraits of a single pulse profile of a pulsar. As a realization of the simulation method, this paper has studied the descattering of Stokes phase portraits of lower frequency observation of PSR B1356$-$60, PSR B1831$-$03, PSR B1859$+$03, PSR B1946$+$35.
\end{abstract}



\section{Introduction}

ISM always cause scintillation and scattering effect to the pulse signals of pulsar (\cite{scheuer68}) and it may damages some of the observing data. The scattering effect of ISM to the pulse profile and the PPA curve have been studied by many authors (\cite{rankin70};\cite{Kuzmin93}; \cite{Bhat03};\cite{Li03};\cite{Abdujappar12}). Some of them have put forward a several useful model to interpret the observing phenomena of pulse broadening and PPA curve flattening (\cite{rankin70}; \cite{komesaroff72}; \cite{wiliamson72}); some of them have studied a model to recover original pulse shape (\cite{Kuzmin93}; \cite{Bhat03}) and PPA curve of some pulsars from scattering effect (\cite{Abdujappar12}).

As mentioned above, the Stokes parameters are disturbed by a scattering effect when the signal travels in the ISM, so it is natural to think that the ISM scattering can affect the Stokes phase portraits of $I$-$Q$, $I$-$U$ and $Q$-$U$. Until now a very few attempts are made to the study of the Stokes phase portraits which hold additional information of the emission geometry of pulsar. Through frequent observation and studying, researchers put up a several models such as the Radius-to-frequency mapping (RFM) model (\cite{lorimer05}) and the Rotating-vector model (RVM) (\cite{Radhakrishnan69}) which were closely related to the geometry of pulsar emission surface. But, recently, researchers (\cite{Chung11} a, b) have analyzed the observing data of some pulsars in a Stokes phase portraits form and they found that the Stokes phase portraits of 24 pulsars have shown that the emission heights inferred from the Stokes tomography technic (\cite{Chung10} a) are different from the emission heights derived from the RFM model. The Stokes phase portraits of known low-latitude emission of 24 pulsars they examined revealed that the 60$\%$ of them may well originated from high altitudes (\cite{Chung10} a). By considering the importance of studying the Stokes phase portraits, we continued the previous work (\cite{Abdujappar12}), in which we mainly examined the descattering of the PPA curve. In this paper, we have noted that ISM scattering can damage and distort some of the observed Stokes phase portraits; for some pulsars, without recovering the data from ISM scattering effect would mislead our analysis (see section 3). We have recovered pulse intensity and PPA curve shape in previous work (\cite{Abdujappar12}) by adding simple tricks to the model of \cite{Kuzmin93}. In this paper, we will study the ISM scattering effect to the Stokes phase portraits of $I$-$Q$, $I$-$U$ and $Q$-$U$ and  will provide the descattering of Stokes phase portraits of four pulsars.  The method is simply described in section 2, results of simulation and application are given in section 3, discussions and conclusions are presented in section 4 and in section 5 respectively.

\section{ A method of compensation for scattering}

According to the distribution scale of irregularities of the electron densities in the ISM, the scattering models are named thin screen, thick screen and extended screen model. The formulae of those models and procedure description of descattering are given below (see \cite{Abdujappar12} for more details).

\begin{equation}
g_{thin}=\exp(-t/\tau_{s}) ~~~~~~~~~~~~~~~~~~~~~~~~~~~~~(t\geq 0)
\end{equation}

\begin{equation}
g_{thick}=(\frac{\pi\tau_{s}}{4t^3})^{1/2}\exp(\frac{-\pi^2\tau_{s}}{16t}) ~~~~~~~~~~~~~~~(t>0)
\end{equation}

\begin{equation}
g_{extend}=(\frac{\pi^5\tau_{s}^3}{8t^5})^{1/2}\exp(\frac{-\pi^2\tau_{s}}{4t}) ~~~~~~~~~~~(t>0)
\end{equation}
where $\tau_{s}$ is scattering broadening time scale which is determined through pulse profile fitting and also from empirical relation between wavelength ($\lambda$) and dispersion measure (DM) (\cite{Ramachandran97}. By using above scattering model, we can hold descattering to the scattered pulse profiles. The frequency spectrum of original pulse $x(t)$ and observed scattered pulse $y(t)$ are related by the equation

\begin{equation}
X(f)=Y(f)/G(f)                                   \\
\end{equation}

\begin{equation}
Y(f)=\int y(t)\exp(-j2\pi ft)dt
\end{equation}

\begin{equation}
G(f)=\int g(t)\exp(-j2\pi ft)dt
\end{equation}
$G(f)$ is the frequency response of ISM, for more details see \cite{Kuzmin93}. The descattered restored total intensity pulse profile is
\begin{equation}
x(t)=\int X(f)\exp(j2\pi ft)df
\end{equation}

The Stokes parameters of $Q$, $U$ are can be derived by assuming $Z(t)$=$Q$+$i$$U$;
the descattered complex number from the frequency spectrum of observed $Z(t)$
 \begin{equation}
 L(f)=Z(f)/G(f)
 \end{equation}
after applying an inverse Fourier transform, we can get
\begin{equation}
L(t)=\int L(f)\exp(j2\pi ft)df;
\end{equation}
from Eq.7 and Eq.9 one can obtain the descattered Stokes parameters $I$, $Q$, $U$. To get the descattered Stokes phase portraits there maybe tow ways to do the best fit. The first one is a fitting the  descattered pulse profile to the intrinsic one, higher frequency signals which have negligible scattering effect; the second one is a iterative method to fit the descattered Stokes phase portraits to the intrinsic one. This paper chosen the first one and has found that the fitting results of scattering time scales of \cite{Abdujappar12} are useful. So, all the fitting results of $\tau_{s}$ of \cite{Abdujappar12} are applied in this paper. In plotting, Stokes parameter $I$ is plotted on the horizontal axis in all figures of Stokes phase portraits of $I$-$Q$ and $I$-$U$ while the $Q$ and $U$ are plotted on the vertical axis. In $Q$-$U$ portraits, $Q$ is plotted on horizontal axis and $U$ is plotted on vertical axis. All plots from F.g1 to Fig.5 are normalized by their own peak value.

\section{Simulation and practical application}
\subsection{Simulation of scattering and descattering of Stokes phase portraits}

The scattering effect and it's descattering for the intensity pulse profiles and PPA curves are simulated in the work of \cite{Abdujappar12}. We have mainly simulated scattering and descattering for Stokes phase portraits of a single pulse profiles of pulsar with each of the three different models. For an illustration of the simulation, the thick screen model is applied here. The simulated pulse profiles have Gaussian shape, with PPA($\psi$) curves following the RVM (\cite{Radhakrishnan69}). This paper has assumed the rate of linear intensity in total intensity is 0.7, so that the Stokes parameters $Q, U$ can be expressed in terms of the linear intensity rate and PPA by $Q=0.7Icos(2\psi)$, $U=0.7Isin(2\psi)$ (\cite{van57}). Then we are able to plot the Stokes phase portraits. The simulated plots are shown in Fig.1; the plots of (a), (b) and (c)are intensity profiles of original, scattered and descattered pulses; the four-panel plots under the plots of a, b, c are PPA curves and Stokes phase portraits of original, scattered and descattered pulses respectively. The each sub-panels of the four-panel plot are PPA curves, $Q$-$U$, $I$-$U$ and $I$-$Q$ phase portraits (clockwise from top left). The simulation shows that ISM scattering causes some sort of changes to the shape of Stokes phase portraits. For instance it causes $Q$-$U$ asymmetry, straitening of $I$-$Q$ (hockey stick shape becomes stick shape, \cite{Chung10} a) and $I$-$U$ shrunk (see Fig.1). In this simulation we take the scattering time scale 35ms and noticed that some of the Stokes phase portraits are not as susceptible as the intensity profiles and PPA curves to the scattering effect. For example $I$-$U$. The scattering effect to the Stokes phase portraits seems small, comparing with the pulse profiles and PPA curve, but it is worth to be noted. It can also found from the figures that, in principle, it is possible to eliminate the scattering effect if the scattering screen is clearly formulated.

 \begin{figure*}[!htb]
   \centering
   \includegraphics[width=3.67cm, angle=-90]{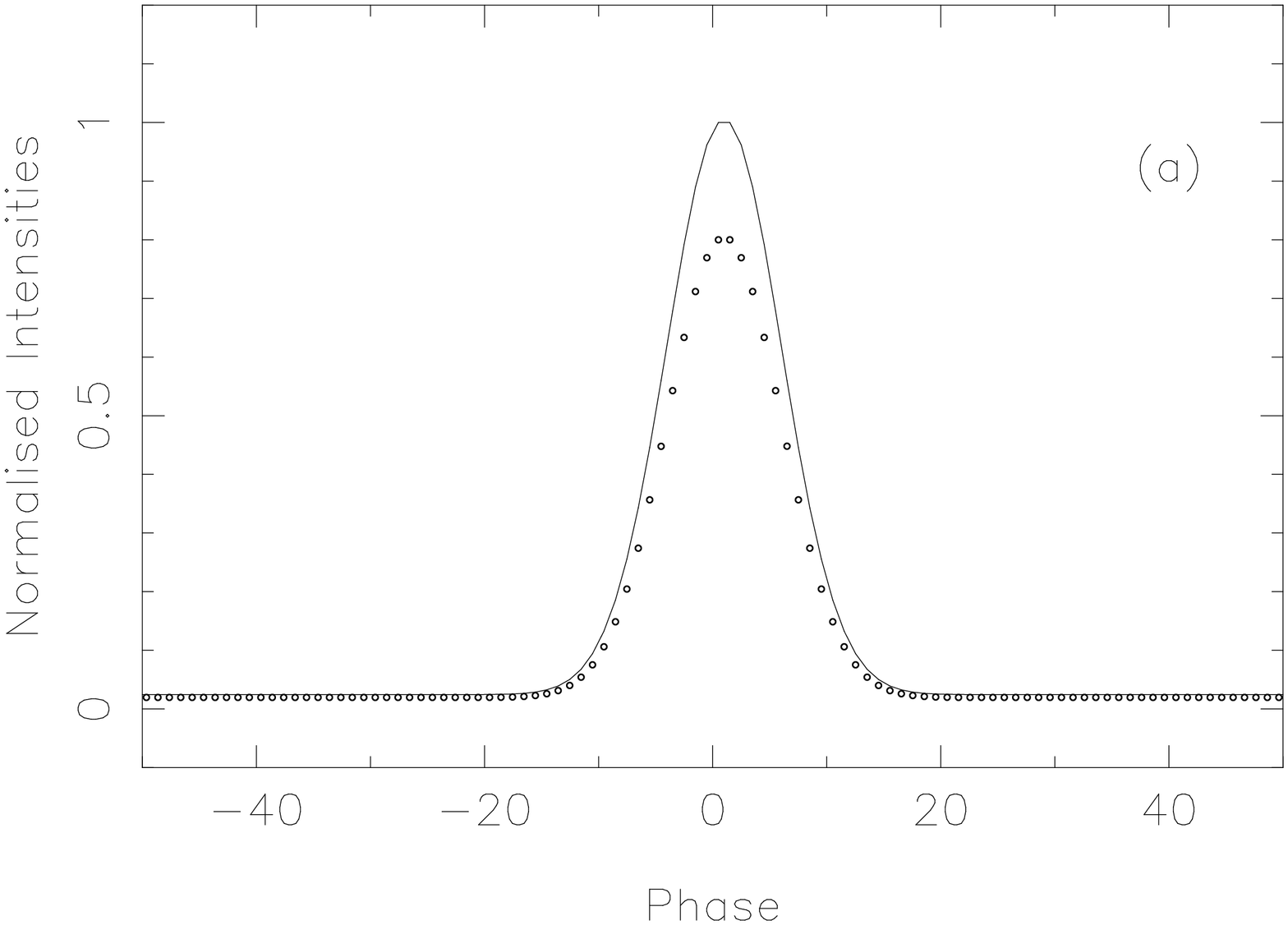}
   \includegraphics[width=3.67cm, angle=-90]{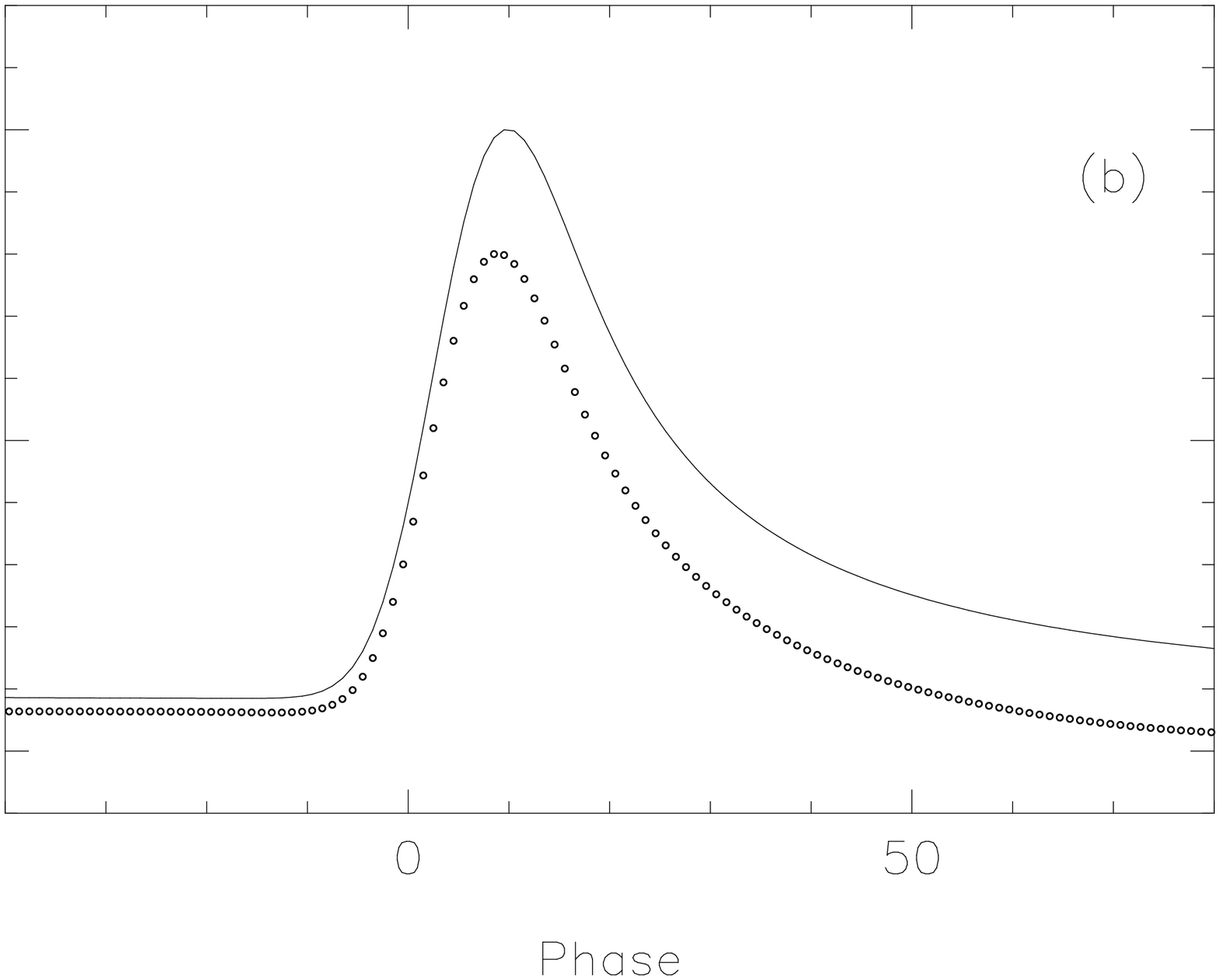}
   \includegraphics[width=3.67cm, angle=-90]{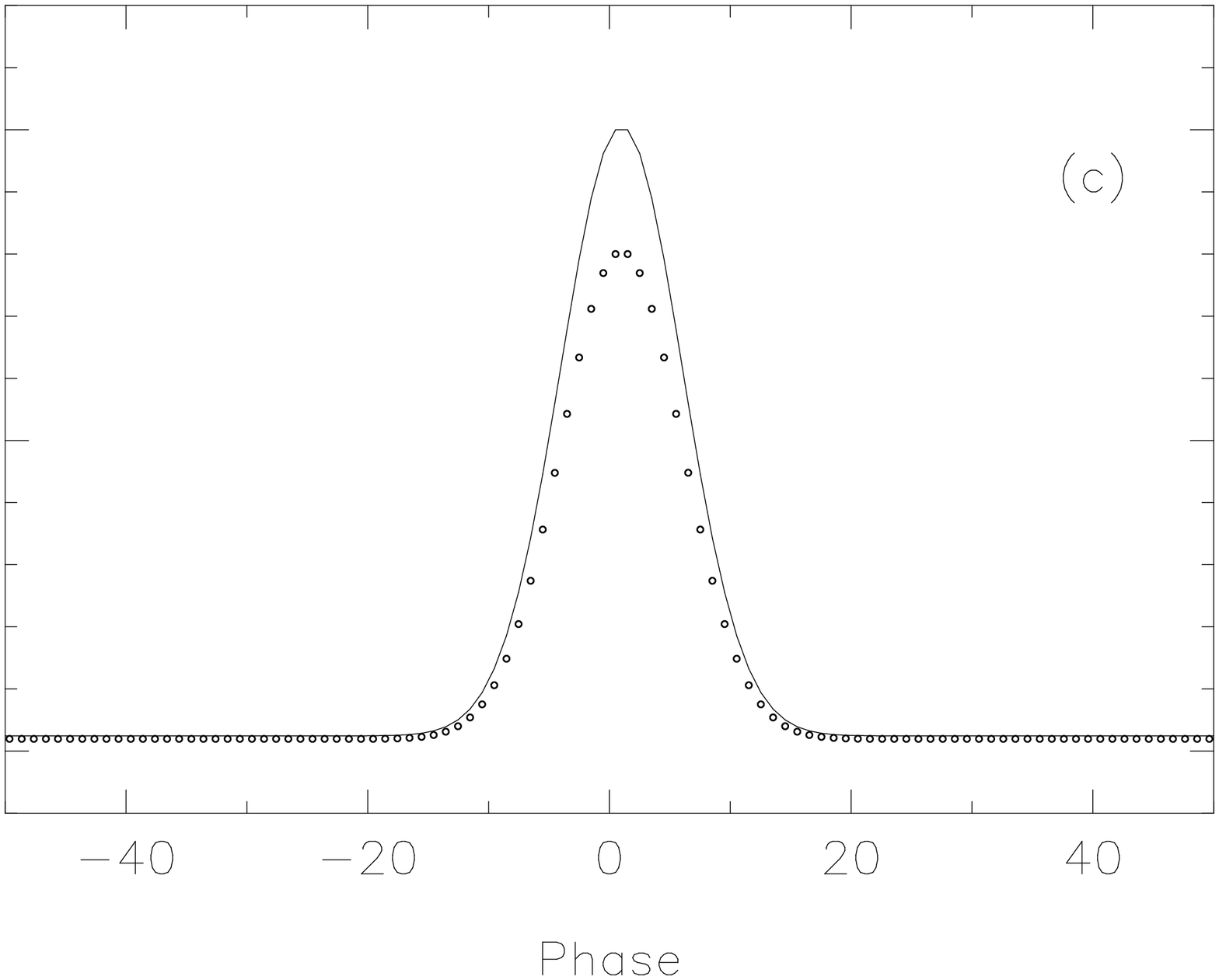}
   \includegraphics[width=3.310cm, angle=-90]{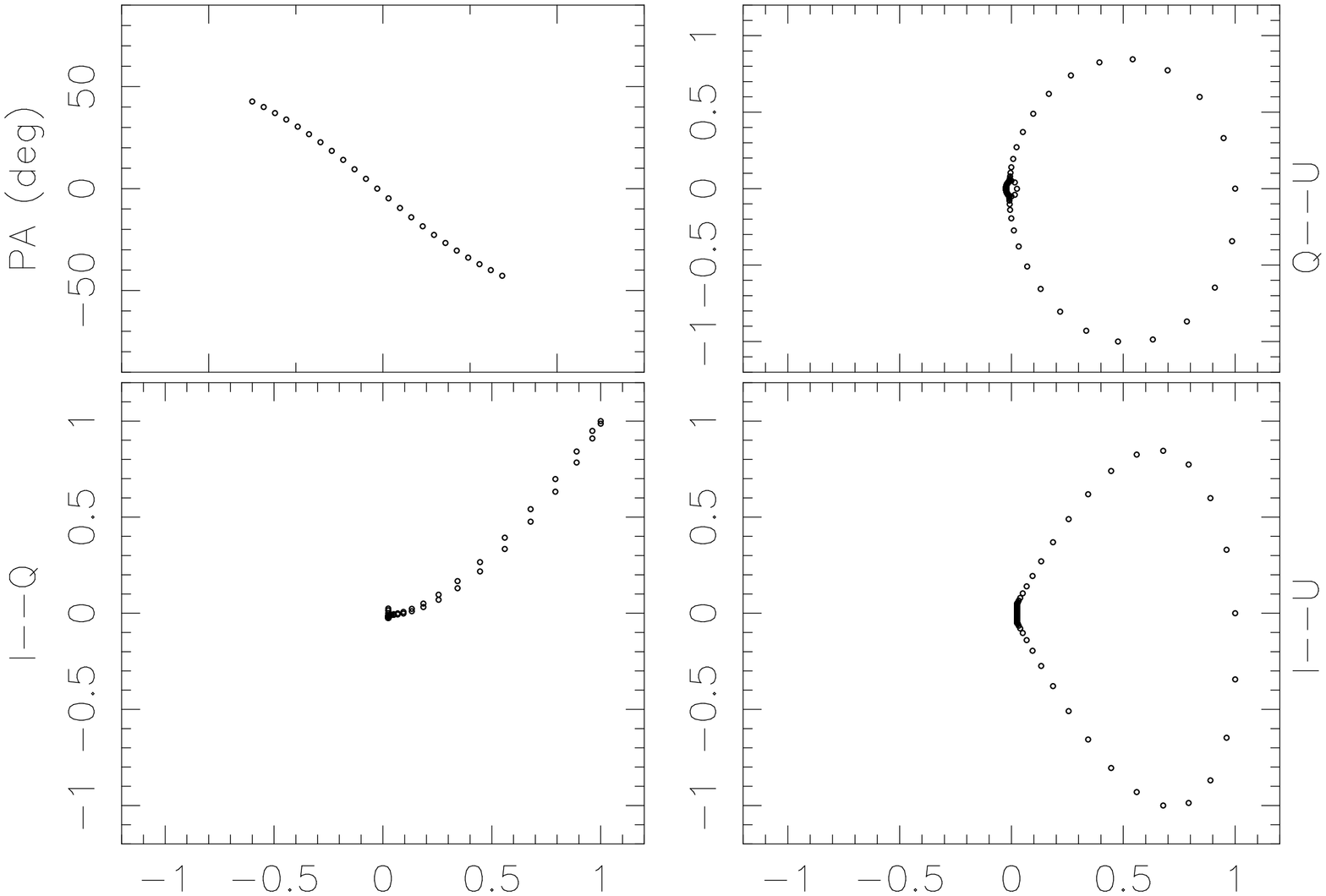}
   \includegraphics[width=3.18cm, angle=-90]{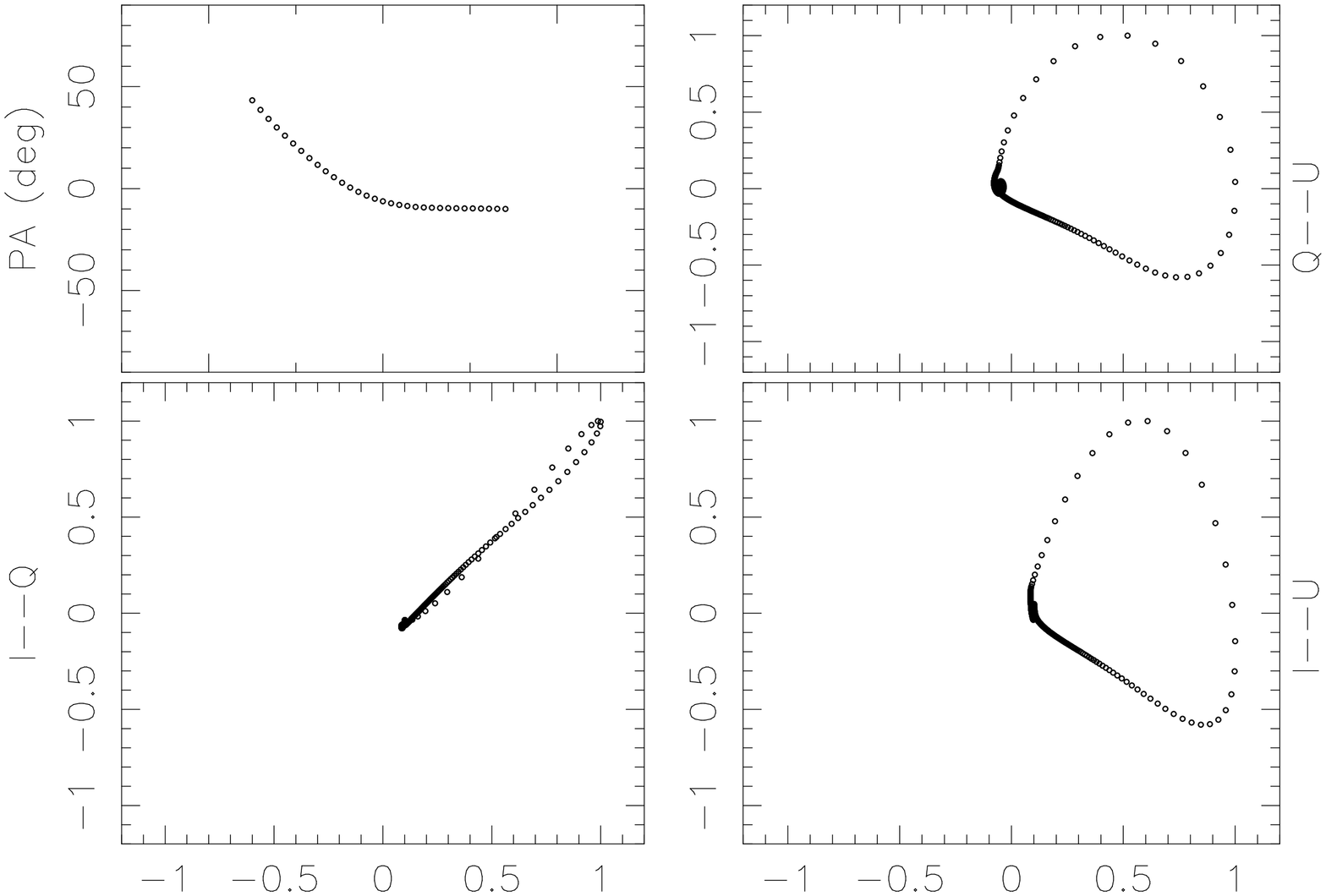}
   \includegraphics[width=3.18cm, angle=-90]{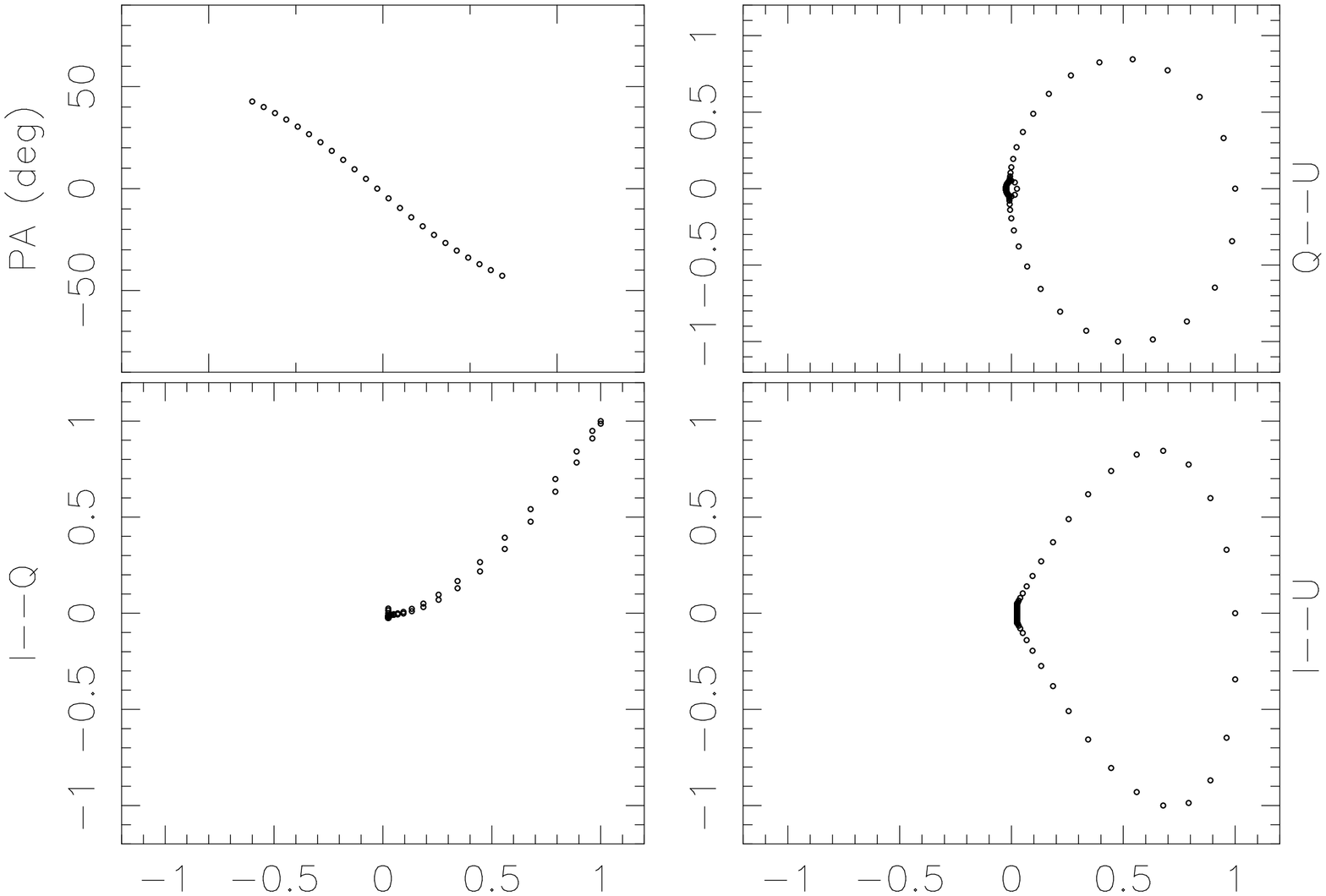}
   \begin{minipage}[]{150mm}

   \caption{ Simulation of Stokes phase portraits of a single pulse for scattering and descattering. the plots of (a), (b) and (c)are intensity profiles of original, scattered and descattered pulses; the four panel plots under the plot of a, b, c are Stokes phase portraits and PPA curves of original, scattered and descattered pulses. the each sub-panels of the four panel plot are PPA curves, $Q$-$U$, $I$-$U$ and $I$-$Q$ (clockwise from top left). } \end{minipage}
   \label{Fig1}
   \end{figure*}

\subsection{Recovering the Stokes phase portraits  }

The data in this paper are downloaded from European Pulsar Network (EPN) online database (\cite{Lorimer98})\footnotemark.\footnotetext{http://www.mpifr-bonn.mpg.de/old$_{-}$mpifr/div/pulsar/data/b\\
rowser.html}  All the observed lower frequency pulse profiles of this four pulsars show obvious pulse profile broadening and PPA curve flattening (see Fig.2-Fi.g.5, top left). A descattering compensation of the intensity pulse profile and PPA curve of the lower-frequency pulses of those pulsars are completed by \cite{Abdujappar12}. The main focus of this section is to present the descattered Stokes phase portraits and its analysis. Observations showed that ISM scattering have no much effect to the high frequency signals than to the lower-frequency one. So the higher frequency pulses's Stokes phase portraits are chosen as an intrinsic Stokes phase portrait (It has no obvious scattering effect in itself) for comparison with the descattered Stokes phase portraits. The scattering time scales of those four pulsars was achieved through fitting the pulse profile in the work of \cite{Abdujappar12}. In this paper , we reviewed previous method and derived the same results for descattering Stokes phase portraits; for brief reference, here we retabulate few of the $\tau_{s}$ and relative data of four pulsars (see table 1).

This paper has tested all $\tau_{s}$ of the three models for obtaining the good descattered Stokes phase portraits. For the convenience of further analysis of the descattered phase portraits, the intensity profiles and PPA curves are given along with Stokes phase portraits in all figures (adopted from \cite{Abdujappar12}). From Fig.2 to Fig.5, the top left four-panel plots are intensity profiles of observed scattered (total intensity profile in (a) and linear intensity profile in (b)), descattered (total intensity profile with thick line in (c) and linear intensity profile with thick line in (d)), and intrinsic signals (total intensity profile with dotted line in (c) and linear intensity profile with dotted line in (d)) respectively (see Fig.2); the observed scattered Stokes phase portraits are placed on the top right panel, the bottom left panel is descattered stokes phase portraits, the bottom right panel is intrinsic Stokes phase portraits. The sub-four-panels of the latter three panels are PPA curves, $Q$-$U$, $I$-$U$, $I$-$Q$ phase portraits (clockwise from top left ). Because, it is hardly possible to construct complete look-up table of Stokes phase portraits by the formula of RVM, so we used the simulated Stokes phase portraits of look-up tables of \cite{Chung10} (a) for analyzing the descattered results. The Stokes parameters of \cite{Chung10} (a) are generated on the base of similar assumption of RVM. They had provided us a large set of Stokes phase portraits with different emission models at different emission height, $\alpha$ and $i$ ($i=\alpha+\beta$); here $\alpha$ is inclination angle between rotational axis and magnetic axis, $i$ observers inclination angel between rotational axis and line of sight, $\beta$ is the impact parameter (\cite{lorimer05}). After descattering, we have tried to check similar phase portraits from the simulated look-up tables (\cite{Chung10} a) to confirm and analyze descattering phase portraits.

\begin{table*}
\centering
\begin{minipage}[h]{146mm}

\caption[h]{ Parameters of four pulsars and their scattering time scales in some scattering models. The parameters are tabulated from column one to column eight as pulsar name, period, dispersion measure, observed higher frequency and lower frequency, empirical value of scattering time scale by Eq. (4), time scale for thick-screen, time scale for extended screen.}\end{minipage}

\small
 \begin{tabular}{ccrrrrrrrrcrl}
  \hline\noalign{\smallskip}
PSR Name & $P$ & $DM$ &  $Freq$  &$Freq$  &$\tau_{em}$    &$\tau_{thick}$  &$\tau_{extend}$ \\
   &$(ms)$  &$(pc$ $cm^{-3})$  &$(GHz)$  &$(GHz)$  &$(ms)$     &$(ms)$   &$(ms)$      \\

  \hline\noalign{\smallskip}
B1356$-$60   &  127.503335  & 294.133   &1.56     &0.659594    &9.88         &0.5   & \\
B1831$-$03   &  686.676816  &  235.800  &0.610    &0.408   &29.63          &     &6.5 \\
B1859$+$03   &  655.445115  & 402.900   &0.925    &0.606   & 60.97        &   &5.5 \\
B1946$+$35   &  717.306765  &  129.050  &0.61    &0.408    &1.87        &6.0     &     \\

  \noalign{\smallskip}\hline
\end{tabular}
\end{table*}

The Fig.1 shows that ISM scattering changed the symmetry of balloon shape of $Q$-$U$ and heart shape of $I$-$U$ and straitened the hockey stick shape of $I$-$Q$. But after descattering, the balloon shape of $Q$-$U$ and heart shape of $I$-$U$ are more symmetric than their scattered one, the scattered stick shape of $I$-$Q$ becomes hockey stick shape again after descattering. The main features of single pulse simulation of phase portraits are balloon and heart shape of $Q$-$U$, heart and banana shape of $I$-$U$ and hockey stick and banana shape of $I$-$Q$ (\cite{Chung10} a). In this simulation, we demonstrated the main effect of ISM scattering to the Stokes phase portraits as an asymmetric shape of $Q$-$U$, $I$-$U$ and a stick shape of $I$-$Q$. The descattered phase portraits of Fig.1 are compared with the Fig.6 of pure dipole field at a low emission altitude of \cite{Chung10} (a). It is found that the original and descattered Stokes phase portraits match well with the plots of Fig.6 to Fig.9 of \cite{Chung10} (a) at $40^\circ$ $\leq$i$\leq$ $50^\circ$ and $10^\circ$ $\leq\alpha\leq$ $20^\circ$, the scattered one does not match any of the Stokes phase portraits in the look-up table. This may implies that the ISM scattering affects the Stokes phase portraits.

Fig.2, PSR B1356$-$60,
shows the descattered Stokes phase portraits of a pulse signal observed at 0.659594 $GH_{Z}$; the higher frequency pulse is observed at 1.56 $GH_{Z}$; all three $\tau_{s}$ are tested, the application of thick screen model is presented here. Unlike the scattered one, the recovered portraits of $Q$-$U$ and $I$-$U$ show more symmetric shape in most of their phase portraits; phase portrait $I$-$Q$ almost remained unchanged. The descattered Stokes phase portraits showed similarity with the Fig.38 to Fig.45 of current-modified dipole field model of \cite{Chung10} (a) at $70^\circ$ $\leq$i$\leq$ $80^\circ$ and $30^\circ$ $\leq\alpha\leq$ $40^\circ$. Descattered phase portraits and intrinsic portraits have similar shape; without considering the direction of Stokes phase portraits the two emission frequency might originated from the same emission height of pulsar surface (\cite{Chung10} a).

 Fig.3, PSR B1831$-$03,
 shows the descattered Stokes phase portraits of a pulse signal observed at 0.4 $GH_{Z}$; the higher frequency pulse is observed at 0.61 $GH_{Z}$; all three $\tau_{s}$ are tested, the application of extended screen model is given here; the descattered Stokes phase portraits $Q$-$U$ and $I$-$U$ showed weak symmetry or regular pattern in their shape. The twisted phase portraits of $I$-$Q$ is recovered as a hockey stick shape (without considering the orientation). The descattered phase portraits found similarity with the Fig.38 to Fig.41 of current-modified dipole field model of \cite{Chung10} (a) at $i\sim 50^\circ$ and $10^\circ$ $\leq\alpha\leq$ $30^\circ$. Although the pulse profiles well matched with the intrinsic one (top left panel in Fig.3 ), but the descattered Stokes phase portraits displayed no similarity with the intrinsic phase portraits.

 \begin{figure*}

   \centering
   \includegraphics[width=4.7cm, angle=-90]{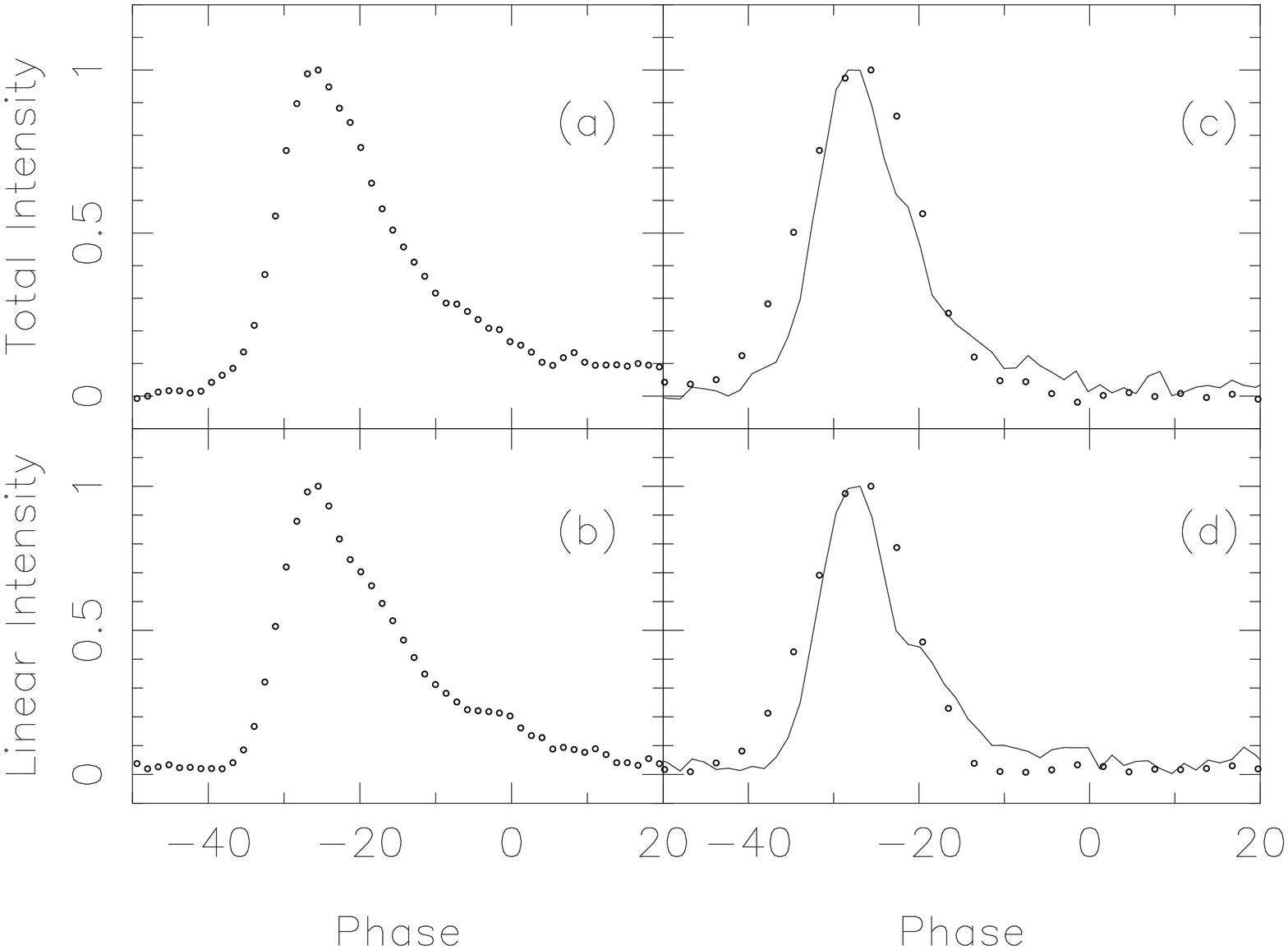}
   \includegraphics[width=4.70cm, angle=-90]{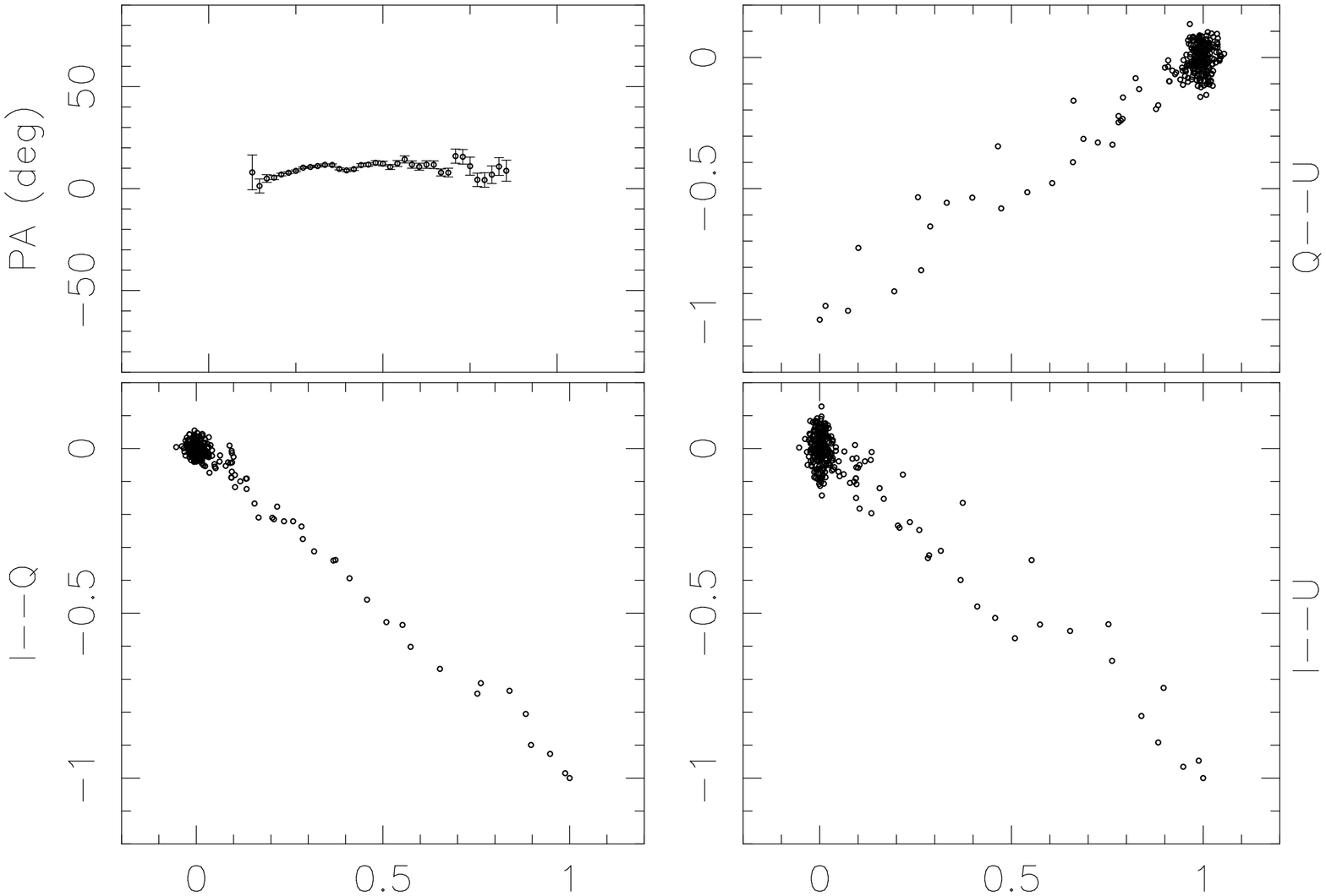}
   \includegraphics[width=4.5cm, angle=-90]{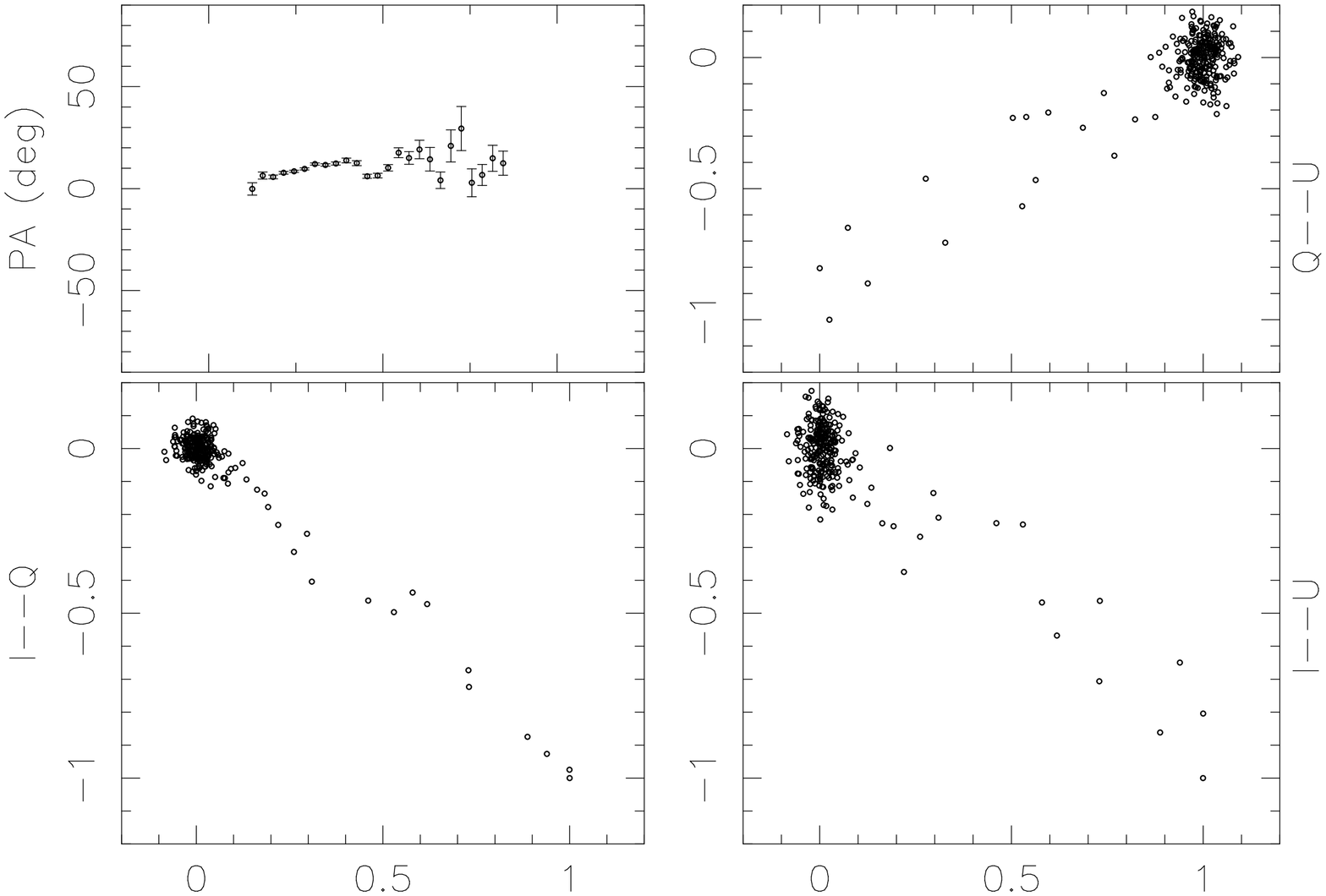}
   \includegraphics[width=4.50cm, angle=-90]{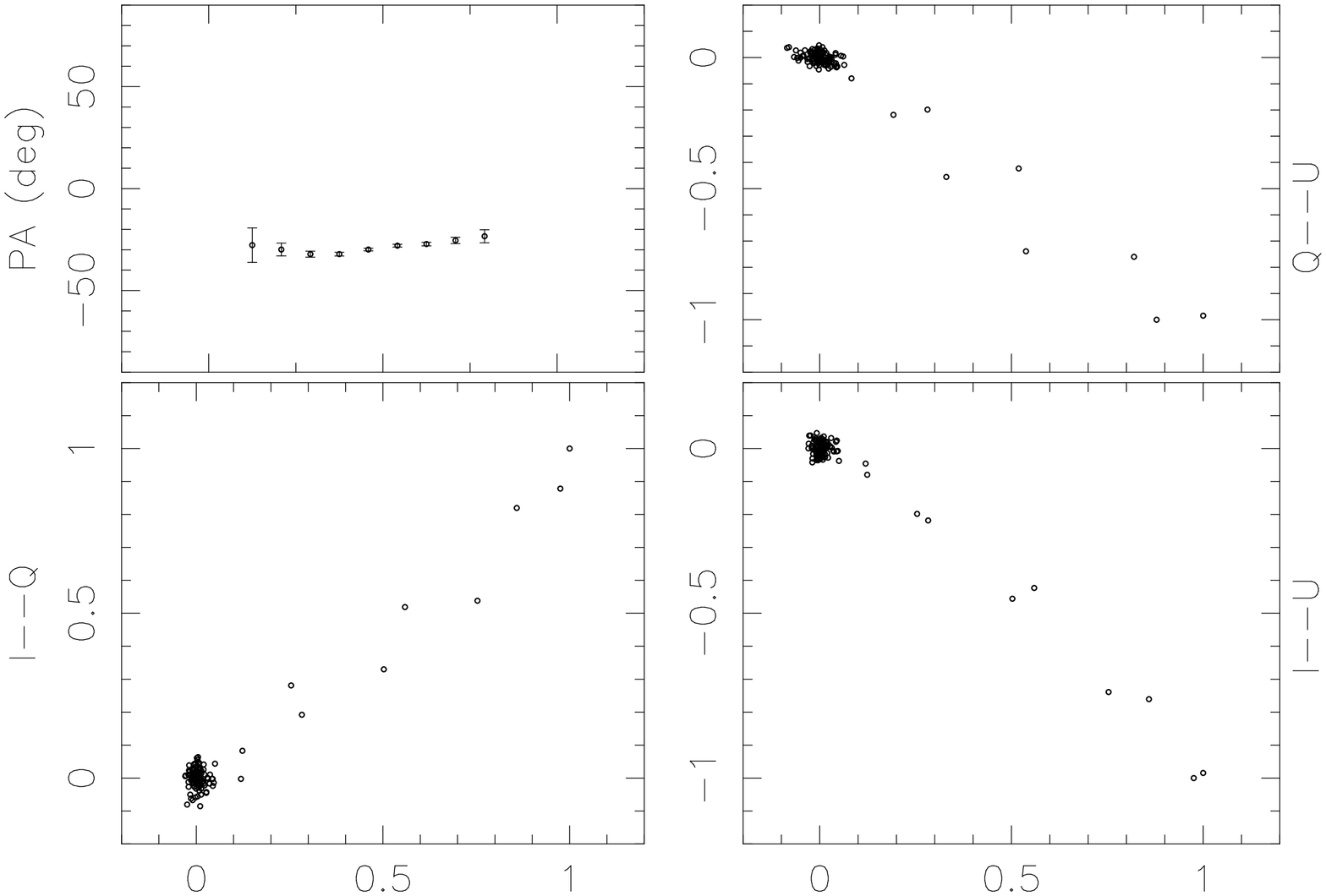}
   \begin{minipage}[]{135mm}
   \caption{  PSR B1356$-$60, the top left four panel plots are observed scattered total intensity profile (a) and linear intensity profile (b). The thick lines in (c) and (d) are descattered total and linear intensity pulse profiles respectively and the dotted lines are observed intrinsic total intensity pulse profile and linear intensity pulse profile. The observed Stokes phase portraits are placed on the top right panel, the bottom left panel is descattered stokes phase portraits, the bottom right panel is intrinsic Stokes phase portraits. The sub-panels of the latter three panels (clockwise from left top) are PPA curves, $Q$-$U$, $I$-$U$, $I$-$Q$.} \end{minipage}

   \end{figure*}

    \begin{figure*}
   \centering
   \includegraphics[width=4.840cm, angle=-90]{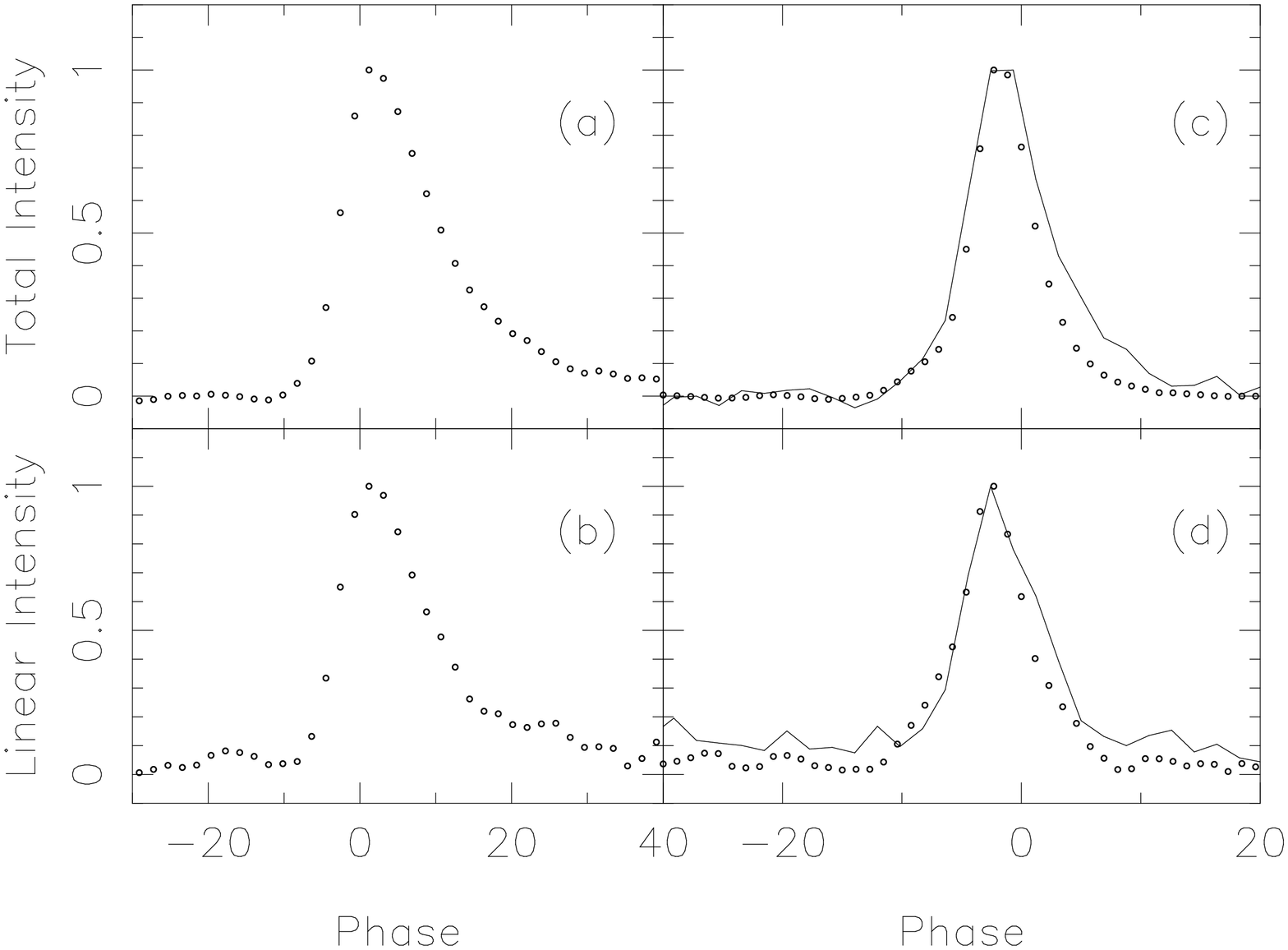}
   \includegraphics[width=4.840cm, angle=-90]{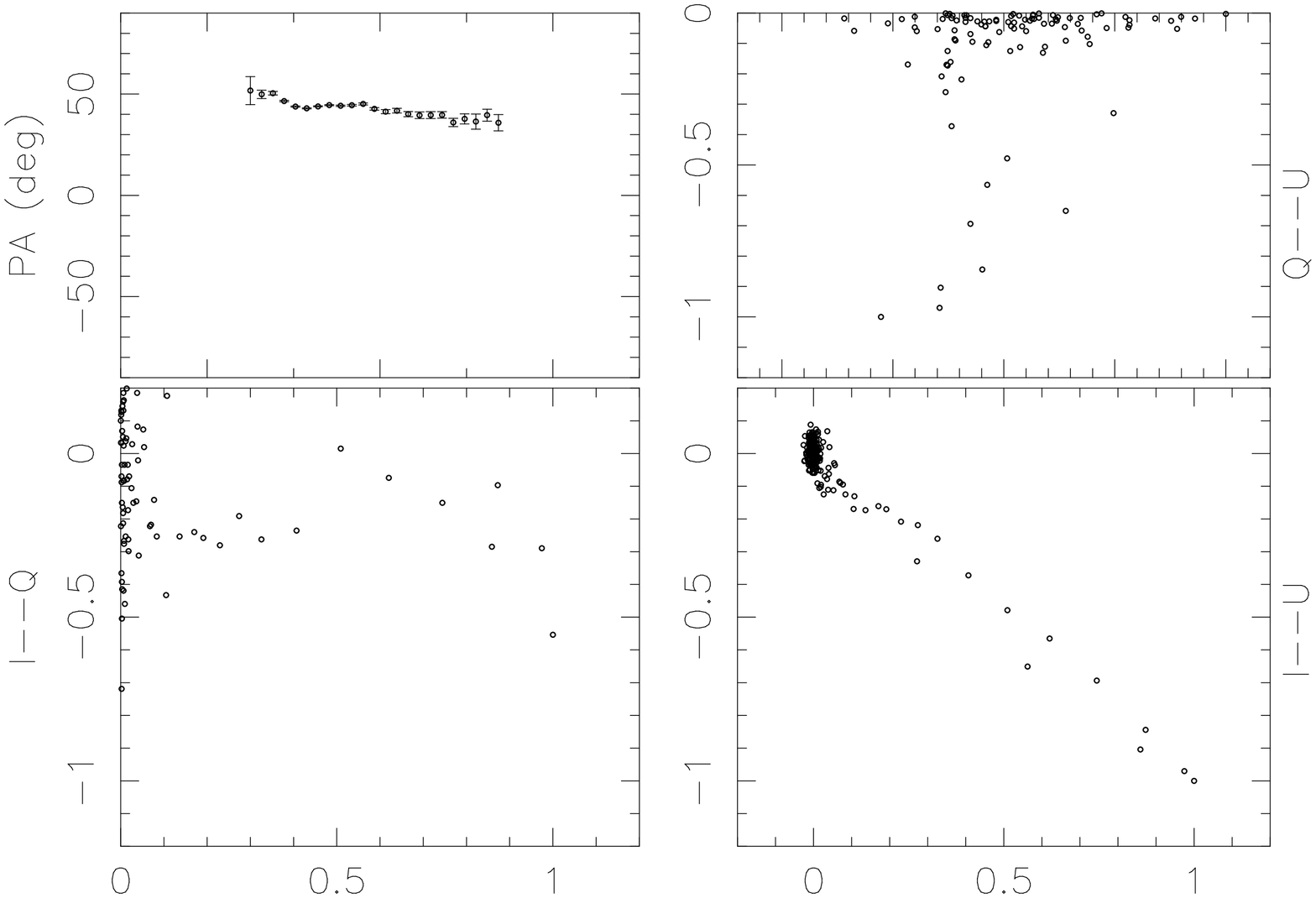}
   \includegraphics[width=4.60cm, angle=-90]{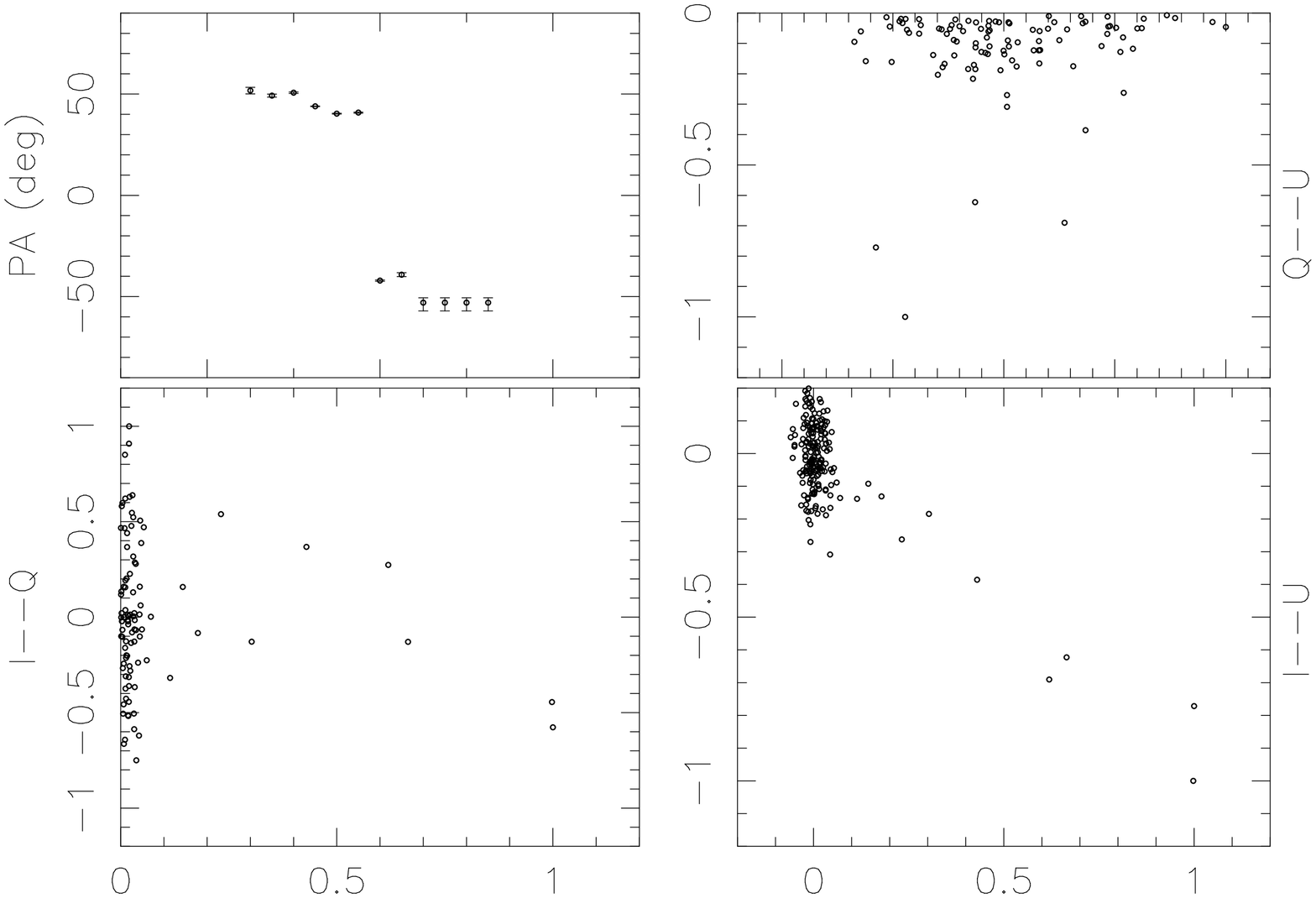}
   \includegraphics[width=4.60cm, angle=-90]{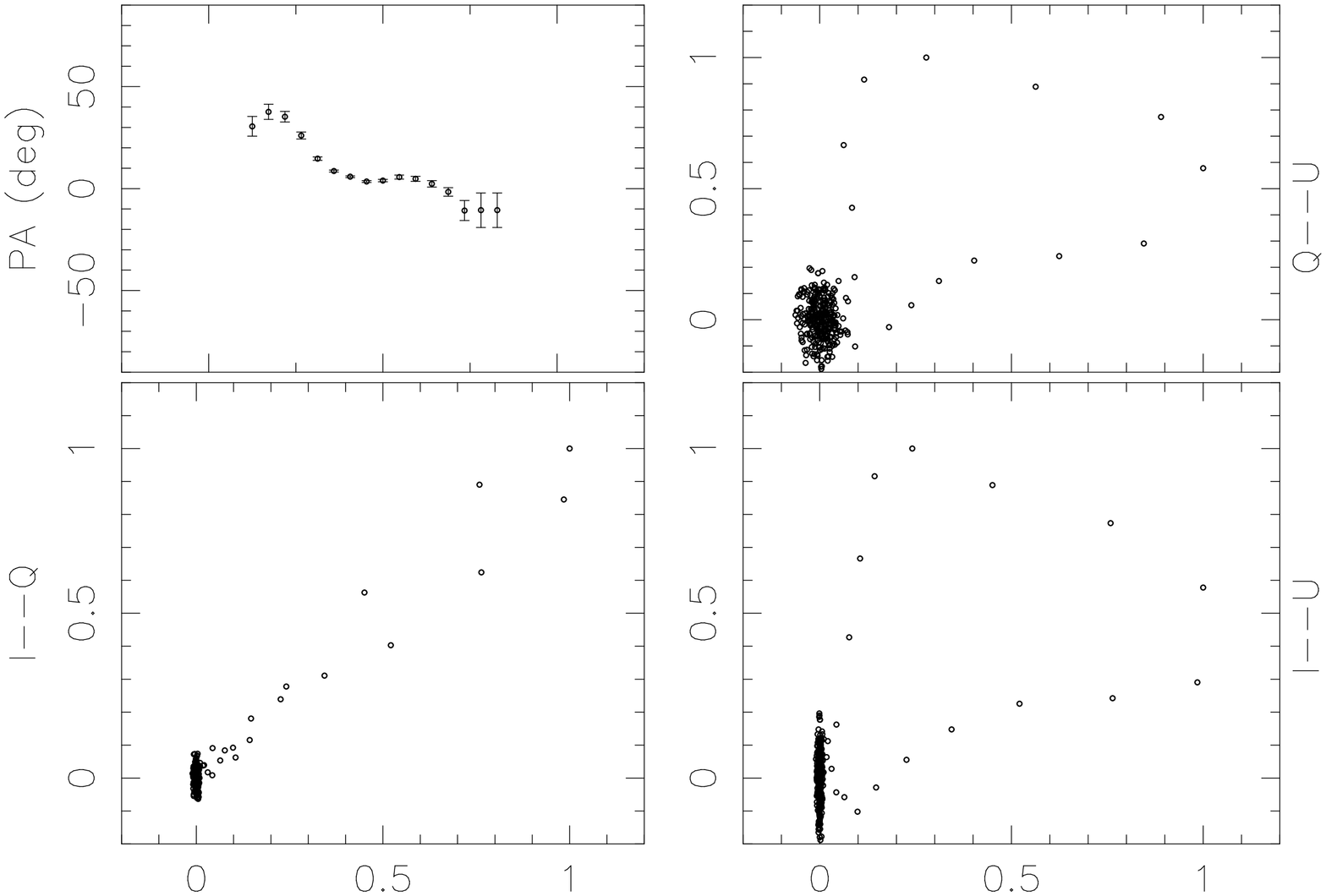}
   \begin{minipage}[]{135mm}

   \caption{ Descattered pulse profiles and Stokes phase portraits for PSR B1831$-$03. See Fig.2 caption for details. } \end{minipage}

   \end{figure*}

Fig.4, PSR B1859$+$03,
presents the descattered Stokes phase portraits of a pulse signal observed at 0.606 $GH_{Z}$; the higher frequency pulse is observed at 0.925 $GH_{Z}$; all three $\tau_{s}$ are tried, the presented plots in Fig.4 are obtained by using extended screen model. The descattered phase portraits $Q$-$U$ showed weak balloon shape, the phase portraits of $I$-$U$ showed symmetric shape. The phase portrait of $I$-$Q$ in scattered and in descattered almost remain it's own shape of banana. The recovered Stokes phase portraits are similar with the Fig.53 to Fig.56 of current-modified dipole field model of \cite{Chung10} (a) at $70^\circ$ $\leq$i$\leq$ $80^\circ$ and $20^\circ$ $\leq\alpha\leq$ $30^\circ$.

Fig.5, PSR B1946$+$35,
 gives the descattered Stokes phase portraits of a pulse signal observed at 0.408 $GH_{Z}$; the higher frequency pulse is observed at 0.61 $GH_{Z}$; all three $\tau_{s}$ are tested, shown here is the application of the thick screen model. The recovered Stokes phase portraits $Q$-$U$, $I$-$Q$ are almost symmetric in their own shape and the $I$-$U$ is banana shape, they are all inflated than the observed scattered one. Not only the direction of all phase portraits of descattered and intrinsic one are similar, but also they have similar phase portraits. For example $I$-$U$, $I$-$Q$. they are probably from the same emission hight (\cite{Chung10} a). The recovered phase portraits showed similarity with the Fig.38 to Fig.45 of current-modified dipole field model of \cite{Chung10} (a) at $10^\circ$ $\leq$i$\leq$ $20^\circ$ and $50^\circ$ $\leq\alpha\leq$ $70^\circ$.

\begin{figure*}
   \centering
   \includegraphics[width=4.750cm, angle=-90]{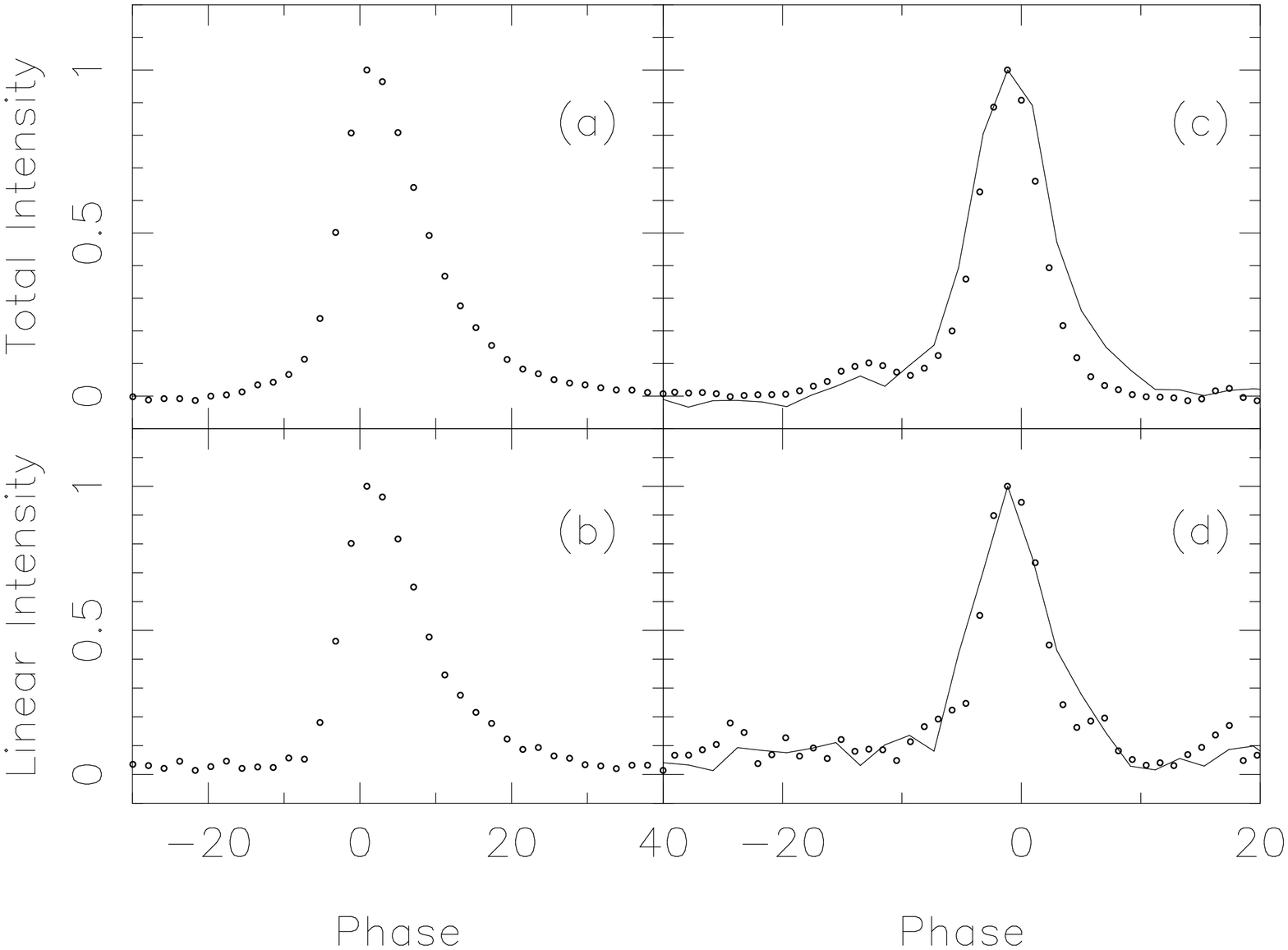}
   \includegraphics[width=4.750cm, angle=-90]{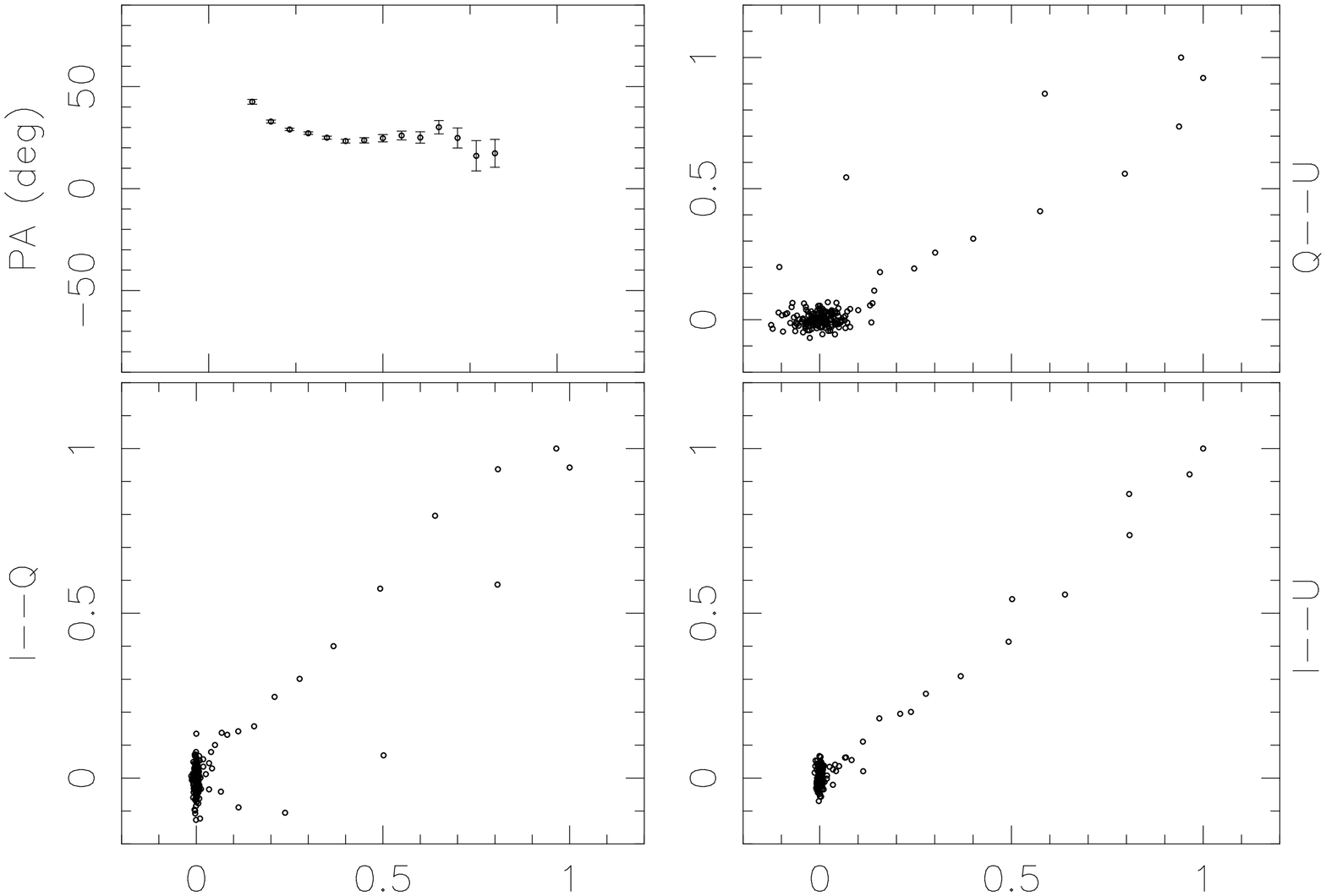}
   \includegraphics[width=4.50cm, angle=-90]{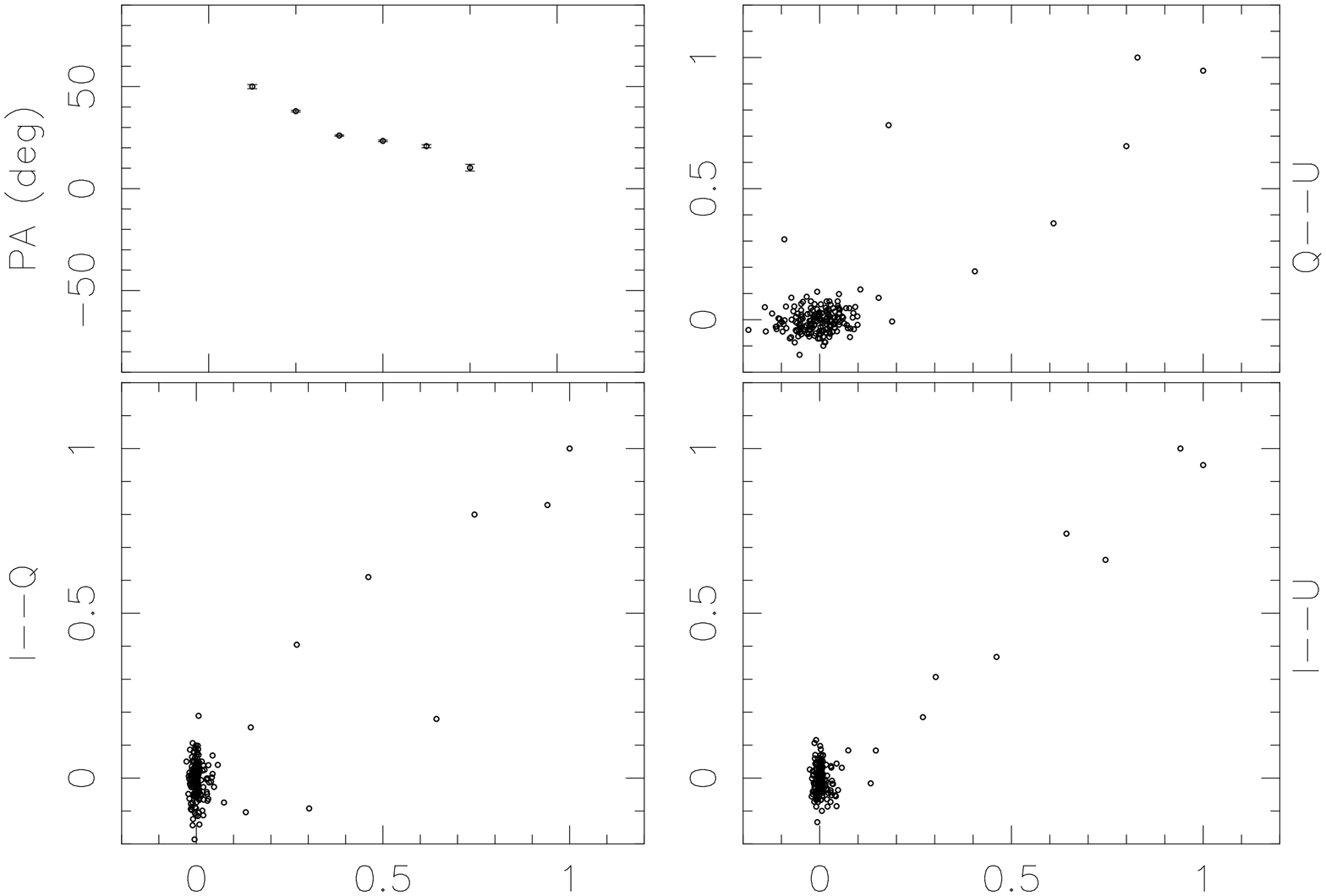}
   \includegraphics[width=4.50cm, angle=-90]{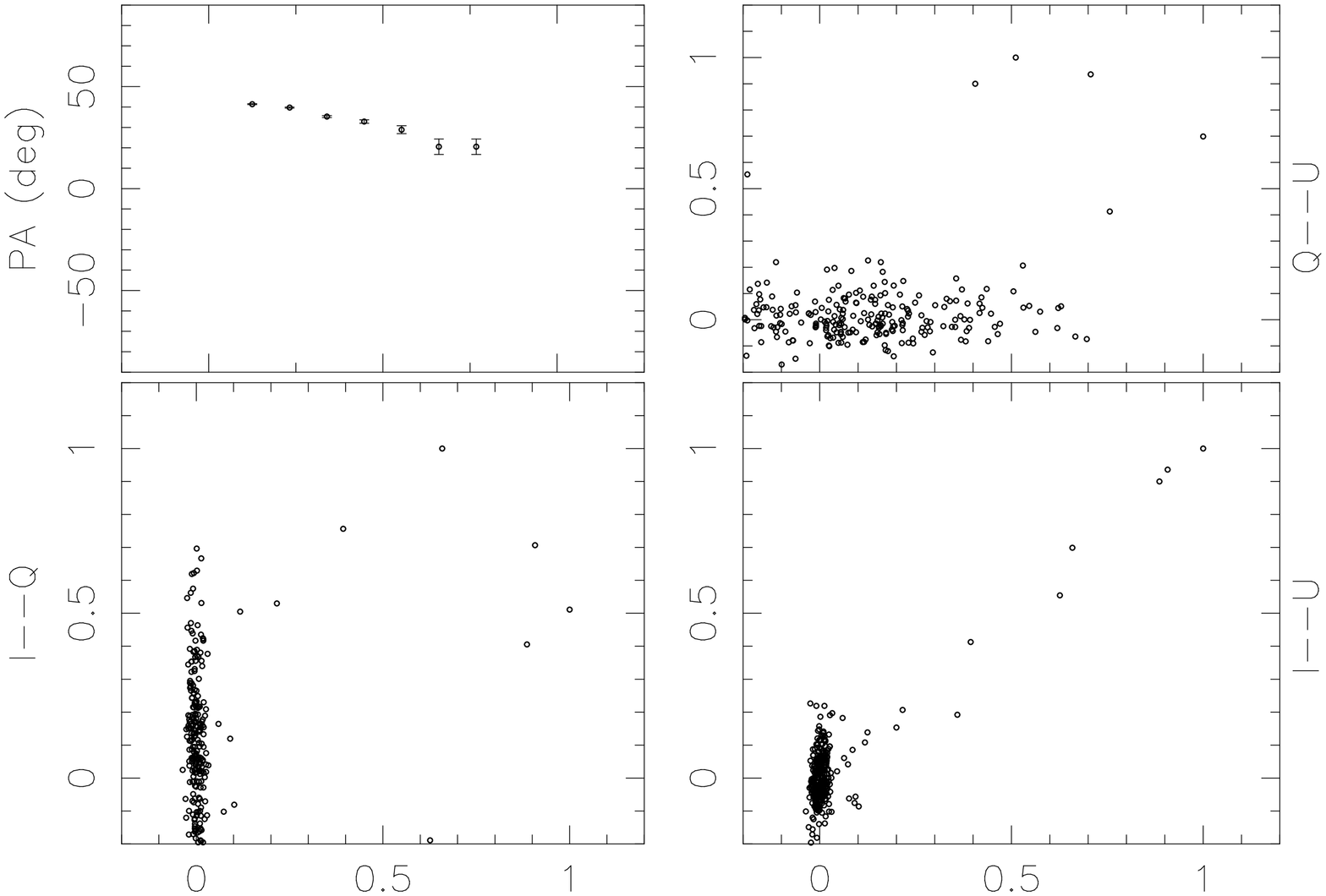}
   \begin{minipage}[]{135mm}

   \caption{ Descattered pulse profiles and Stokes phase portraits for PSR B1859$+$03. See Fig.2 caption for details.  } \end{minipage}

   \end{figure*}

\begin{figure*}
   \centering
   \includegraphics[width=4.750cm, angle=-90]{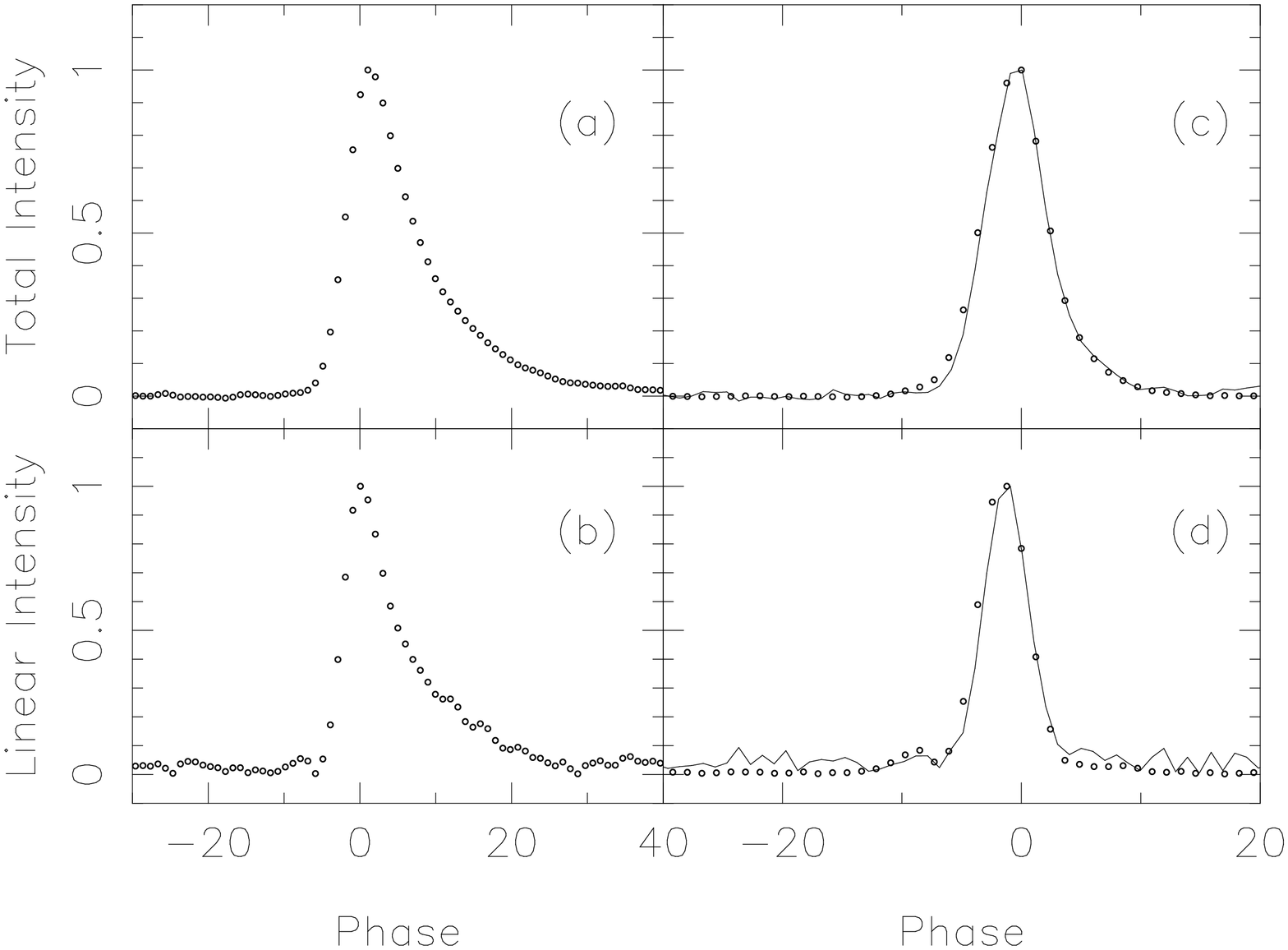}
   \includegraphics[width=4.750cm, angle=-90]{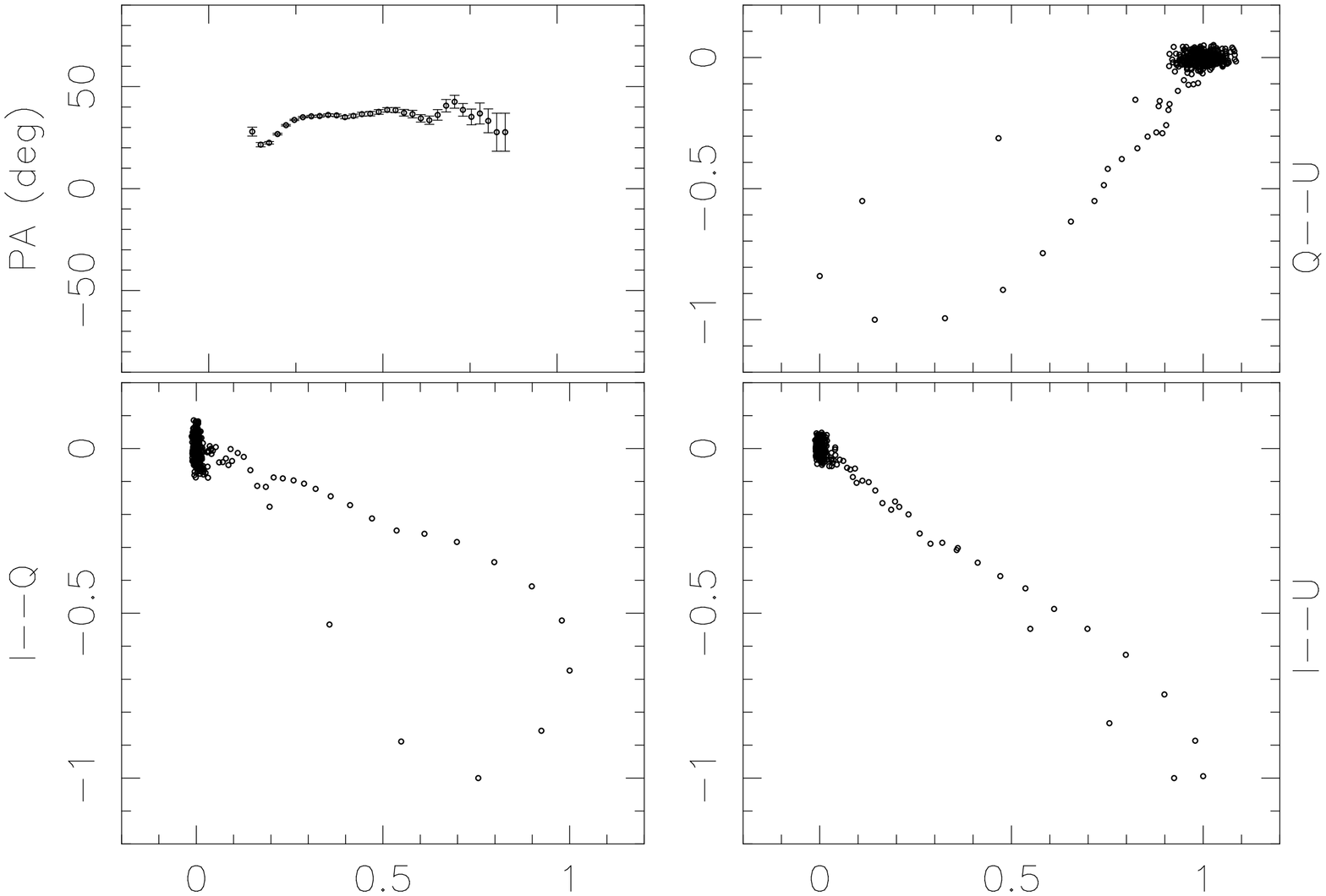}
   \includegraphics[width=4.50cm, angle=-90]{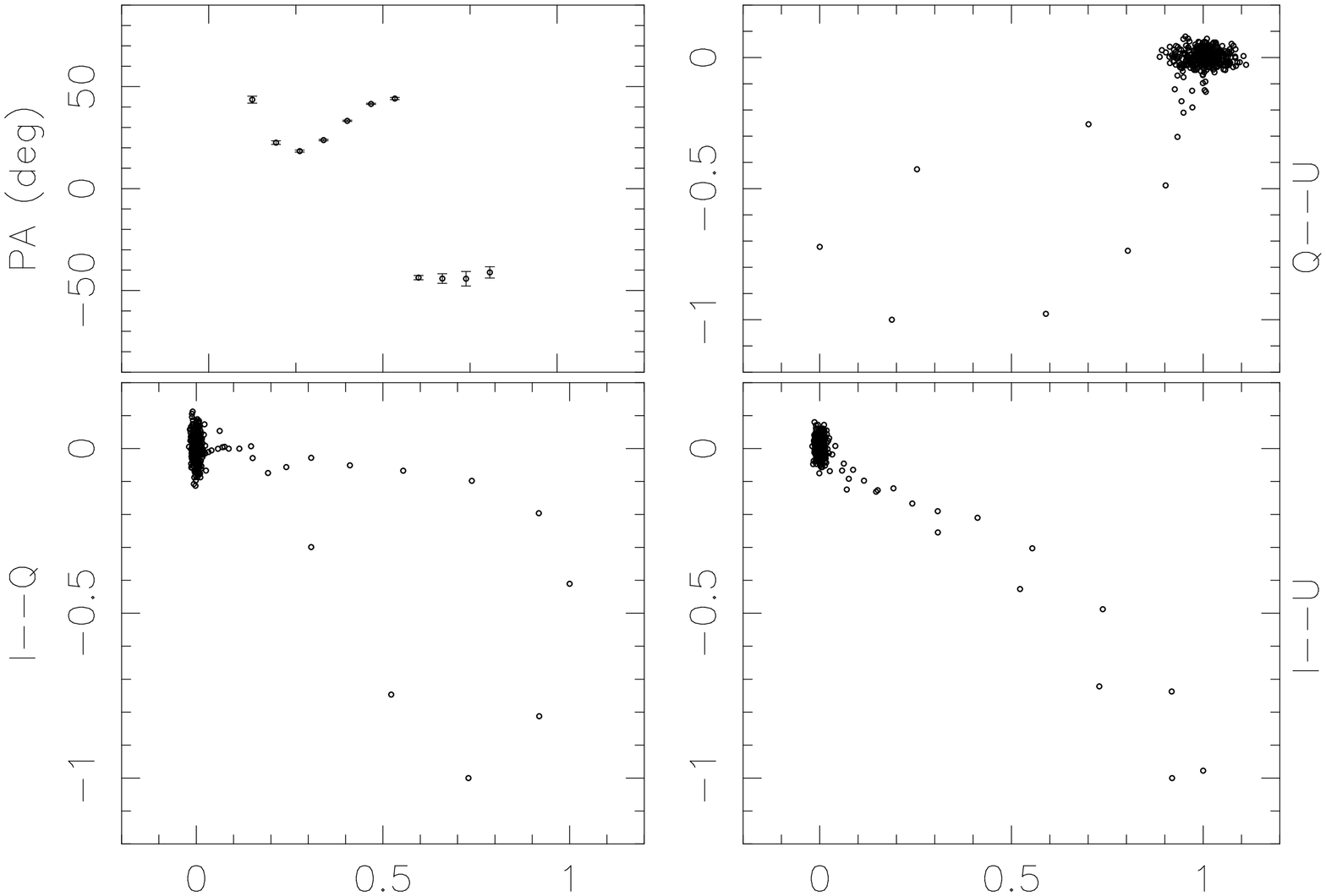}
   \includegraphics[width=4.50cm, angle=-90]{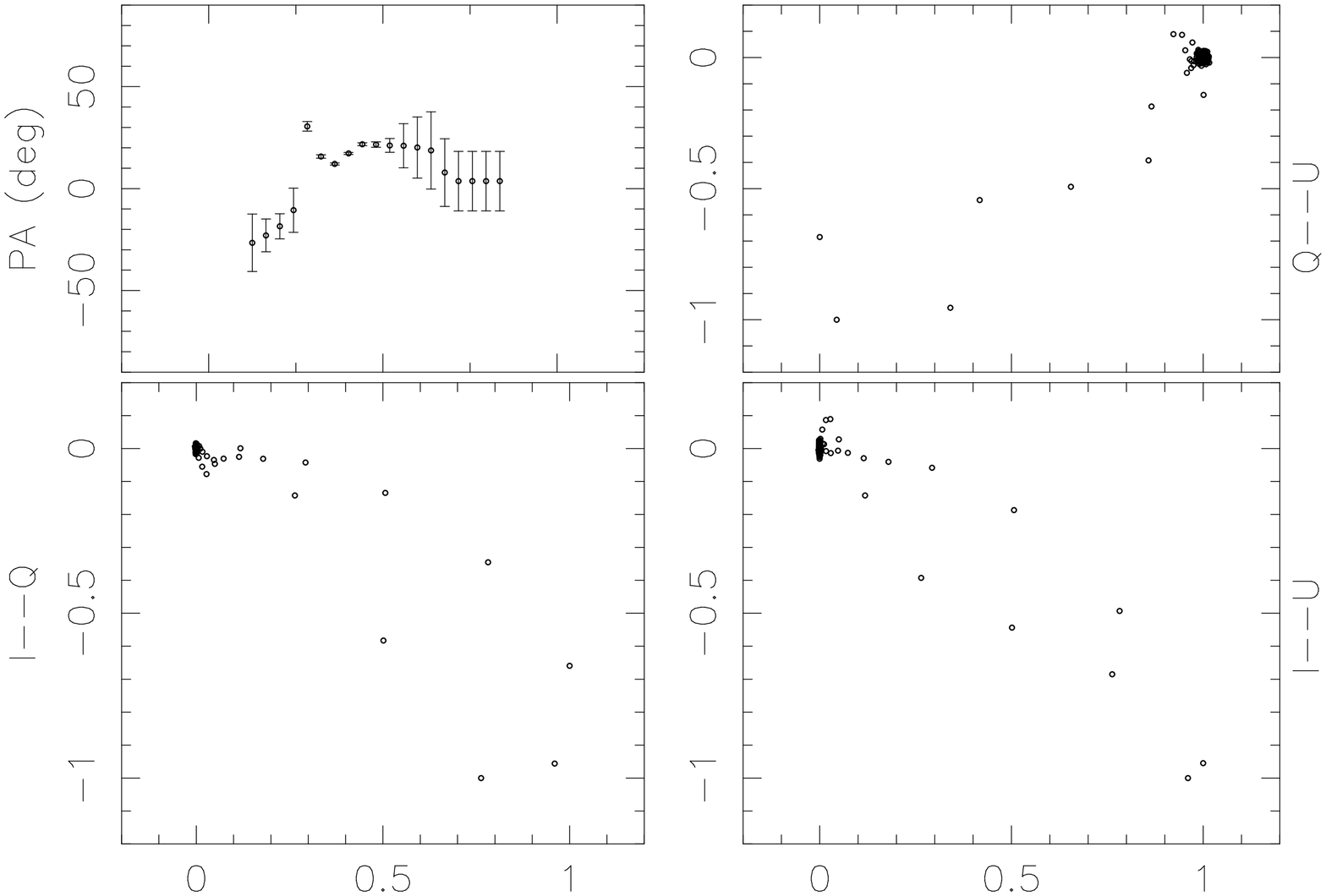}
   \begin{minipage}[]{135mm}

   \caption{Descattered pulse profiles and Stokes phase portraits for PSR B1946$+$35. See Fig.2 caption for details.  } \end{minipage}

   \end{figure*}

\section{Discussion}

In section 3.1, we have held a simulation to show a ISM scattering and descattering.  Simulation showed the basic features of Stokes phase portraits of a single pulse profile, such as symmetric balloon shape of $Q$-$U$, heart shape of $I$-$U$ and hockey stick shape of $I$-$Q$, and showed the main effects of ISM scattering to the shape of Stokes phase portraits, for example, asymmetric shape of $Q$-$U$, $I$-$U$ and stick shape of $I$-$Q$. The descattered results of the simulation bring the possibility of recovering the original shape of Stokes phase portrait. The scattering phenomena not only causes pulse profile broadening and PPA curve flattening, but also it causes a distortion to the Stokes phase portraits, for example, the clear distorted shape of $Q$-$U$ and $I$-$U$ of PSR B1356-60, $Q$-$U$ and $I$-$Q$ of PSR B1831-03, $Q$-$U$ of PSR B1859+03 and $I$-$U$ of PSR B1946+35. Of course, in general, the irregularity and distortion of phase portraits may be related to the magnetosphere configuration. In section 3.2, after descattering the scattered data of four pulsars, the phase portraits of $Q$-$U$ and $I$-$U$ of PSR B1356-60, $Q$-$U$ and $I$-$U$ of PSR B1831-03, $Q$-$U$ and $I$-$U$ of PSR B1859+03 and $Q$-$U$ and $I$-$U$ of PSR B1946+35 showed a regular shape, symmetric balloon and banana shape, the phase portraits of $I$-$Q$ of PSR B1831-03, $I$-$Q$ of PSR B1859+03 and $I$-$Q$ of PSR B1946-35 showed symmetric shape, hockey stick and banana shape. It is noted that the recovered shape of $Q$-$U$, $I$-$U$ and $I$-$Q$ manifested the features of Stokes phase portraits which are simulated in this paper and in \cite{Chung10} (a).

When analyzing the descattered phase portraits we have used the look-up tables of Stokes phase portraits which are provided by \cite{Chung10} (a). It is found that, without removing the scattering effect from the simulated and the observed scattering data, we can not find resemblance of the Stokes phase portraits from the look-up tables (\cite{Chung10} a). This implies that it is necessary to hold descattering compensation to some scattered data. We also have noted that not only the descattered phase portraits well matched with the simulated data of \cite{Chung10} (a), but also the PPA curves are matched too. For PSR B1859$+$03, we used the simulated Stokes phase portraits of hollow cone model with $\mid \alpha-i\mid \sim50$ (\cite{Chung10} a); this indicates that the hollow cone pulse profile is single peaked and it is the same with the single peak pulse profile of PSR B1859$+$03. The large values of $\mid \alpha-i\mid$ of four pulsars can be associated with their shallow PPA swing curve; on the other hand, when the beam is cut at its outer edge, the impact angle $\mid \alpha-i\mid$ is large \cite{lorimer05}.

 The pulse width of associated Stokes phase portraits of \cite{Chung10} (a) are all the same as 10$^\circ$. The pulse width of the four pulsars in this paper are calculated by $W=2.5^\circ P^{-0.5}$ (\cite{krz12}) and we obtained the pulse width as 7$^\circ$, 3$^\circ$, 3$^\circ$, and 3$^\circ$ degree respectively. So, the range of values of $i$ and $\alpha$ for the four pulsars refereed above are used as a reference for analyzing the descattered Stokes phase portraits, they need further study to be confirmed. As \cite{Chung10} (a) said the Stokes phase portraits should be analyzed concurrently with the PA swing and pulse profiles in future when interpreting radio pulsar polarization data.

\section{Conclution}

In this paper we have presented a simulation of scattered and descattered Stokes phase portraits of a single pulse. And we have shown the descattering compensation of Stokes phase portraits of four pulsars. We have noticed from the Fig.1-Fig.5 that ISM scattering, in some extent, distorts the Stokes phase portraits and we found the descattering method implemented in this paper not only worked well to the scattered intensity profiles and PPA curves (\cite{Abdujappar12}), but also useful to the Stokes phase portraits of the observed scattered pulse signals. Descattering the Stokes phase portraits allows us to obtain the original shapes of Stokes phase portraits of scattered pulse signals. It is worth to noted that it is necessary to consider the effect of scattering to the observed Stokes parameters of lower frequency signals, while more attentions are given to the effect of rotational aberration, current-flow (\cite{Johann A01}) and rotation and modulation (\cite{Dyks08}).

\acknowledgments
This work was funded by the
National Natural Science Foundation of China (NSFC) under No. 11273051 and the key program project of Joint Fund of Astronomy by NSFC and CAS under No. 11178001.

\end{document}